\documentclass[runningheads]{llncs}
\usepackage[T1]{fontenc}
\usepackage{graphicx,verbatim}
\usepackage{amsmath} 
\usepackage{booktabs}
\usepackage{enumitem}
\usepackage{wrapfig}
\usepackage[pagebackref=false,breaklinks=true,colorlinks=true,bookmarks=false,citecolor=blue,urlcolor=blue,linkcolor=blue]{hyperref}
\usepackage{booktabs}      
\usepackage{colortbl}      
\usepackage{xcolor}        
\usepackage{multirow}      
\usepackage{array}         
\definecolor{gold}{RGB}{255,215,0}
\definecolor{silver}{RGB}{192,192,192}
\definecolor{bronze}{RGB}{205,127,50}
\usepackage{float}
\usepackage{floatflt}
\usepackage{graphicx}
\usepackage{amssymb}

\begin{document}
\title{EndoPerfect: High-Accuracy Monocular Depth Estimation and 3D Reconstruction for Endoscopic Surgery via NeRF-Stereo Fusion}
%
\author{Pengcheng Chen*\inst{1} \and
Wenhao Li*\inst{2,3} \and
Nicole Gunderson\inst{1} \and
Jeremy Ruthberg\inst{4} \and
Randall Bly\inst{4} \and
Zhenglong Sun\inst{2,3} \and
Waleed M. Abuzeid\inst{4} \and
Eric J. Seibel\inst{1}
}

\authorrunning{P. Chen, W. Li et al.}

\institute{
University of Washington, Mechanical Engineering, Seattle WA 98195, USA \and
The Chinese University of Hong Kong, Shenzhen, China \and
AIRS, Computer Science and Engineering, Shenzhen, China \and
University of Washington, Otolaryngology, Seattle WA 98195, USA \\
\email{\{pengcc,nmgundo,eseibel\}@uw.edu, \{jruthb,randbly,wabuzeid\}@uw.edu,}\\
\email{wenhaoli1@link.cuhk.edu.cn, sunzhenglong@cuhk.edu.cn}
}

\maketitle              
\begin{abstract}
In endoscopic sinus surgery (ESS), intraoperative CT (iCT) offers valuable intraoperative assessment but is constrained by slow deployment and radiation exposure, limiting its clinical utility. Endoscope-based monocular 3D reconstruction is a promising alternative; however, existing techniques often struggle to achieve the submillimeter precision required for dense reconstruction. In this work, we propose an iterative online learning approach that leverages Neural Radiance Fields (NeRF) as an intermediate representation, enabling monocular depth estimation and 3D reconstruction without relying on prior medical data. Our method attains a point-to-point accuracy below 0.5 mm, with a demonstrated theoretical depth accuracy of 0.125 ± 0.443 mm. We validate our approach across synthetic, phantom, and real endoscopic scenarios, confirming its high accuracy and reliability. These results underscore the potential of our pipeline as an iCT alternative, meeting the demanding submillimeter accuracy standards required in ESS.


\end{abstract}

\section{Introduction}
Approximately 250,000 Endoscopic Sinus Surgery (ESS), procedures are performed annually in the United States \cite{bhattacharyya2010ambulatory}. ESS is commonly performed on patients with chronic rhinosinusitis (CRS) refractory to non-surgical therapy  \cite{orlandi2016international}. This surgery involves operating near several critical anatomical structures, including the cranial olfactory nerves, the dura mater, the orbits, and the carotid artery, presenting a risk of potentially catastrophic injury. The damage to critical structures is estimated to be between 1\% and 3\% \cite{stankiewicz2011complications}. Additionally, due to the complexity of the surgery and insufficient information about surgical completeness, 15-30\% of ESS surgery cases necessitate a second revision surgery, which is mainly caused by incomplete surgical dissection resulting in persistent inflammatory disease, leading to ongoing symptoms in CRS patients \cite{baban2020radiological}. Generally, surgeons rely on intraoperative computed tomography (iCT) to measure the completeness of surgery. A proof-of-concept study demonstrated that iCT revealed residual bony partitions in 30\% of ESS cases, particularly in the frontal or ethmoid sinuses, which altered the surgical plan and drove immediate additional surgical intervention \cite{jackman2008use}. 

\begin{figure}
    \centering
    \includegraphics[width=0.9\linewidth]{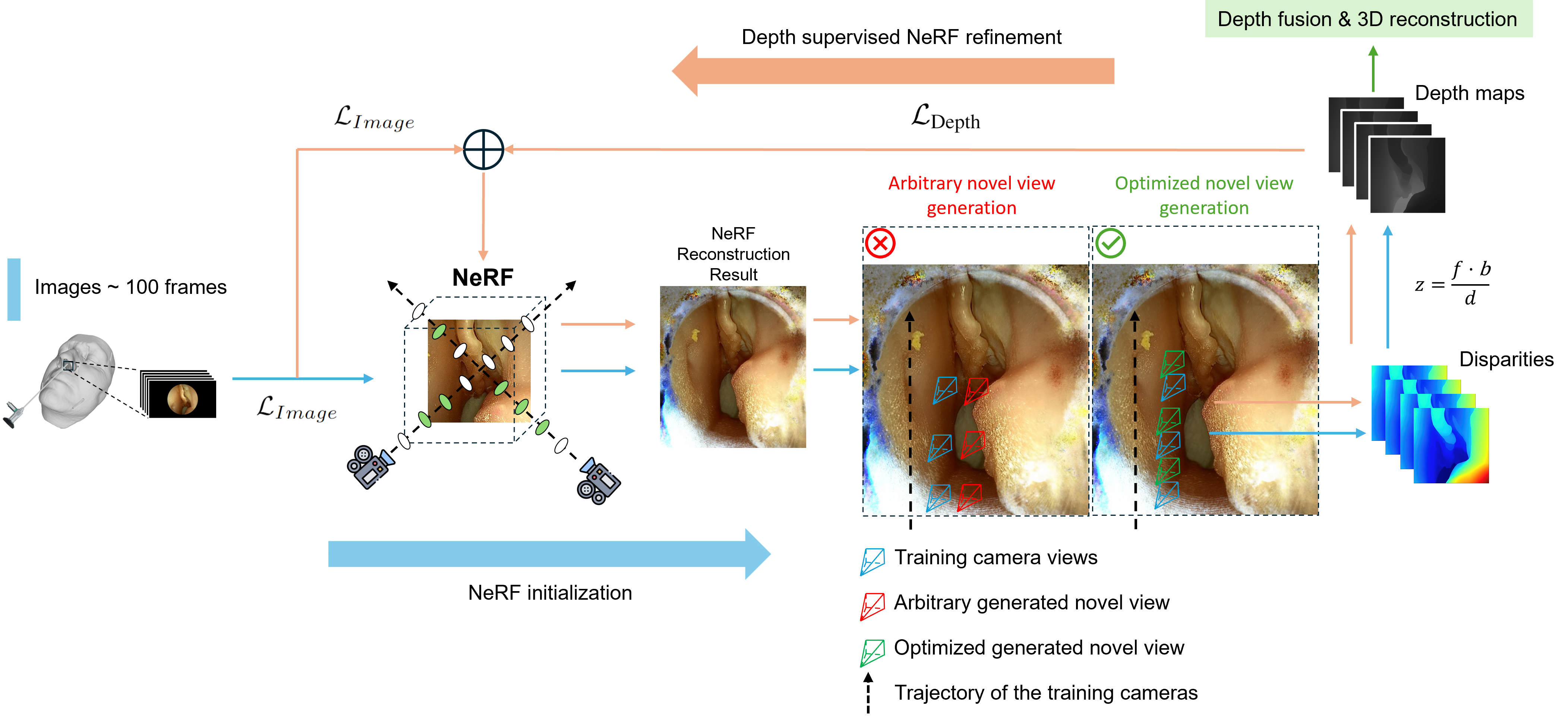}
    \caption{Pipeline overview: First, an initial scene is reconstructed using NeRF. Next, optimized novel stereo views are generated within this scene for stereo depth estimation. The resulting depth maps are then used to supervise the subsequent NeRF training round. This iterative process continues until the depths converge, after which the final depth maps are fused for 3D reconstruction.}
    \label{fig:pipeline}
\end{figure}

Despite the substantial assistance that iCT provides in surgery, with its information significantly influencing the surgical plan, the use of iCT remains uncommon in ESS procedures, as it extends the surgical time thereby elevating intraoperative risk and adding enormous economic cost \cite{schroder2023intraoperative}. Moreover, the substantial radiation exposure inherent to iCT poses health risks to the patient and surgeon alike. Therefore, providing a more convenient, faster, radiation-free method with accuracy comparable to iCT for detecting structural changes during surgery would be highly valuable.

Endoscope-based 3D reconstruction offers an intuitive, cost-effective alternative to iCT; to be a useful substitute of iCT, it requires: \textbf{1) High accuracy:} Submillimeter point-to-point accuracy post-scale recovery for anatomical fidelity; \textbf{2) Efficiency:} Complete pipeline within 15 minutes to match/exceed conventional iCT throughput; \textbf{3) Robustness:} Consistent quality across diverse patients and phenomena.

Current approaches include: \textbf{1) MVS/SLAM-based methods:} Feature-based techniques \cite{chen2022real,lurie20173D} slow in dense 3D reconstruction($\geq$ 30 min), while faster SLAM methods compromise accuracy; \textbf{2) Monocular depth estimation:} Methods like EndoSfmLearner \cite{ozyoruk2021endoslam}, AF-SfmLearner \cite{shao2022self}, and large visual models (SurgicalDino \cite{beilei2024surgical}, EndoDAC \cite{cui2024endodac}) offer faster reconstruction but limited point-to-point accuracy in ESS due to the domain gap; \textbf{3) Neural rendering:} NeRF or 3D Gaussian splitting \cite{kerbl20233d} to generate dense 3d reconstruction but struggle with mesh quality. In ESS, Liu et al.'s monocular approach \cite{liu2020reconstructing,liu2019dense} delivered state-of-art(SoTA) results, but recent evaluations show significant registration errors averaging >6mm \cite{mangulabnan2023quantitative}.

To solve the problem mentioned above, we propose a novel pipeline using NeRF as an intermediary with binocular depth estimation. This method includes two main parts: NeRF initialization and depth-supervised NeRF iteration; the process is shown in Fig. \ref{fig:pipeline}. Our approach generates stereo views via NeRF, obtains depth maps through binocular estimation, applies iterative NeRF refinement, and fuses these maps into a dense reconstruction. By training NeRF online without prior medical data, we ensure consistent performance across scenarios while bypassing the slow sparse-to-dense process of MVS methods, achieving intraoperative CT replacement-level accuracy (<0.5mm) with improved speed.

\section{Method}
\subsubsection{NeRF initialization:}

In this work, the first step is to obtain a coarse 3D scene representation by initializing with NeRF.  NeRF takes a set of input images \( I = \{I_1, I_2, I_3, \ldots\} \) along with their 5D camera poses, which consist of spatial coordinates \( (p_x, p_y, p_z) \) and viewing directions \( (\theta, \phi) \). It then outputs volume density and radiance, enabling the reconstruction of high-fidelity scenes via volume rendering \cite{10.1145/3503250}. To achieve a fast and efficient initialization, we employ Nerfacto \cite{10.1145/3588432.3591516} as the initialization method for NeRF. This workflow provides a foundation for our pipeline, which consists of three key stages:

\begin{enumerate}
\item \textbf{Hash Encoding:}
Camera positions $\mathbf{p}=(p_x,p_y,p_z)$ are encoded via Hash Encoding~\cite{muller2022instant},
\begin{equation}
h(x) \;=\;\bigoplus_{i=1}^{d} x_i \,\pi_i \;\bmod\; N,
\label{eq:hash_function}
\end{equation}
where $\bigoplus$ is bit-wise XOR, $\pi_i$ are large primes, and $N$ is the hash table size. The encoded positions are fed into a multilayer perceptron $\text{MLP}_{1}$ to produce density $\sigma$ and feature $\zeta(\mathbf{p})$.

\item \textbf{Spherical Harmonics Encoding:}
Camera directions $(\theta,\phi)$ are encoded using Spherical Harmonics. The feature $\zeta(\mathbf{p})$ at position $\mathbf{p}$ is $\zeta(\mathbf{p})\in \mathbb{R}^{|\mathbf{k}_{:-1}|}$ where $\mathbf{k} = \{k_\ell^m : -\ell\leq m \leq \ell, 0 \leq \ell \leq \ell_{\text{max}}\},$
and $\mathbf{k}$ represents the spherical harmonics coefficients. The view-dependent feature $\gamma(\mathbf{d})$ is 
\begin{equation}
\gamma(\mathbf{d}) \;=\;
\bigl[Y_{0}^{0}(\mathbf{d}),\,Y_{1}^{-1}(\mathbf{d}),\,\dots,\,
Y_{\ell_{\max}}^{m}(\mathbf{d})\bigr].
\label{eq:querying_Spherical_Harmonics_Encoding_functions}
\end{equation}
\item \textbf{NeRF and Rendering:}
The encoded view-dependent feature $\gamma(\mathbf{d})$, $\zeta(\mathbf{p})$, and appearance embedding $\ell_i^{(a)}$ are concatenated and input into another separate $\text{MLP}_2$ to produce the final RGB via volume rendering:
\begin{equation}
T=\sum_i \tau_i\bigl(1-\exp\bigl(-\sigma_i \delta_i\bigr)\bigr)\,c_i,\quad
\tau_i=\exp\Bigl(-\sum_{j<i}\sigma_j \delta_j\Bigr),
\label{eq:Render_fuction}
\end{equation}
where $\sigma_i$ is the density, $\delta_i$ is the distance between consecutive samples, and
\begin{equation}
c_i(t) \;=\;
\text{MLP}_2\bigl(\zeta(t),\,\gamma(\mathbf{d}),\,\ell^{(a)}_i\bigr).
\label{eq:color}
\end{equation}
The loss function is $\mathcal{L}_{\text{Image}} \;=\;\|I - I^*\|^2_2,$
where $I$ is the original RGB image and $I^*$ is the rendered image. This loss is used for backpropagation to optimize the MLP parameters, refining the final NeRF reconstruction.
\end{enumerate}



\subsubsection{Novel Stereo Views Generation:}

After generating the initial NeRF scene, we create novel stereo views inside the trained NeRF scene for stereo depth estimation. Since the endoscope's motion is restricted to a small region, random novel view placement causes poor rendering results and NeRF artifacts, degrading depth accuracy. We address this using NeRF's differentiability with gradient descent to optimize stereo view placement.

Our gradient-based approach enforces three geometric constraints: (1) novel stereo views coplanar with corresponding training cameras to maintain epipolar geometry, (2) fixed baseline distance for consistent depth estimation, and (3) unchanged camera orientation for robust stereo matching. We employ mean Zero-mean Normalized Cross-Correlation (ZNCC) as the similarity metric in our optimization objective to maximize matching quality.

\begin{equation}
T_{\text{novel}}^{*}
= 
\arg\max_{T_{\text{novel}} \in \mathcal{C}}
\;\;
\overline{\text{ZNCC}}\bigl(I_{\text{train}},\, I_{\text{novel}}(T_{\text{novel}})\bigr),
\end{equation}
\begin{equation}
\mathcal{C} 
= 
\Bigl\{
  T_{\text{novel}}
  \,\Big|\,
  p_{z}^{\text{novel}} = p_{z}^{\text{train}}, \;
  \|\mathbf{p}_{xy}^{\text{novel}} - \mathbf{p}_{xy}^{\text{train}}\|
     = \text{b}, \;
  \mathbf{R}_{\text{novel},z} = \mathbf{R}_{\text{train},z}
\Bigr\}.
\end{equation}
After obtaining the optimized novel stereo views, we perform disparity estimation on each training-novel stereo pair, followed by depth computation using the standard binocular triangulation formula, where $f$ is the focal length, $b$ is the baseline distance between the stereo pair, and $\mathcal{D}(\cdot)$ denotes the disparity estimation function:

\begin{equation}
   d = \frac{f b}{\mathcal{D}\bigl(I_{\text{train}},\, I_{\text{novel}}\bigr)}
\end{equation}

\subsubsection{Iterative Refinement:}

After obtained the depth maps via stereo depth estimation, we leverage them to supervise NeRF reconstruction, thereby enhancing the quality of NeRF reconstruction. These depth maps are then used for depth supervision, where the depth loss follows DS-NeRF \cite{deng2022depth}. For each image \(I_j\) and its camera pose \(\mathbf{p}_j\), we estimate the depth of visible keypoints \(z_{i}\) by simply projecting \(z_{i}\) with \(\mathbf{p}_j\),taking the re-projected z value as the keypoint’s depth \(d_{ij}\). Depth estimation errors are handled by modeling \(z'_{ij}\) as a Gaussian distribution: \(z'_{ij} \sim \mathcal{N}(z'_{ij}, \hat{\sigma}_i)\). The depth supervision loss \(\mathcal{L}_{\text{Depth}}\) is derived by minimizing KL divergence:

\begin{equation}
\mathcal{L}_{\text{Depth}} = \mathbb{E}_{z_i \in Z_j} \left[ \sum_{k} \log h_k \exp \left(- \frac{(t_k - d'_{ij})^2}{2\hat{\sigma}_i^2} \right) \Delta t_k \right]
\label{eq: Depth Loss}
\end{equation}

With the depth supervision, the overall loss is: \(\mathcal{L} = \mathcal{L}_{\text{Image}} + \lambda_d \mathcal{L}_{\text{Depth}}\). We then iteratively generate stereo views, perform stereo depth estimation, and apply depth supervision to refine the NeRF reconstruction. The process continues until convergence between consecutive iterations, where the depth maps from the final iteration are denoted as \(D^{\text{final}} = \{d^{\text{final}}_1, d^{\text{final}}_2, d^{\text{final}}_3, \ldots, d^{\text{final}}_n\}\). The convergence criterion is defined as \(\|1 - |D^{\text{final}}/D^{\text{final-1}}|\| < \varepsilon\), yielding an optimized depth map set \(D^{\text{final}}\).

To minimize depth estimation errors, we leverage the relationship between depth uncertainty and baseline length: \(\Delta d = \frac{d^2}{f \cdot b} \cdot \Delta _\text{disparity}\). We progressively increase the baseline during NeRF iterations according to \(b_{\text{new}} = b + \lambda_{\text{baseline}} \cdot N_{\text{Iteration}}\). The final depth maps \({D^{\text{final}}}\) are then refined through ray casting and geometric consistency optimization with camera poses to generate geometrically consistent fused depth maps.

\begin{table}[tp]
\centering
\footnotesize
\resizebox{0.67\textwidth}{!}{
\begin{tabular}{@{}lcccccc@{}}
    \toprule
    Direct Methods& Source & MAE (mm) $\downarrow$ & RMSE (mm) $\downarrow$ & STD (mm) $\downarrow$ & $\delta < 1.1$ $\uparrow$ & $\delta_{1.25}$ $\uparrow$ \\
    \midrule
    DepthAnythingV2 (ViT-L)\cite{yang2024depth} & CVPR 2024 & 3.021 & 3.695 & \cellcolor{bronze!25}2.035 & 9.58\% & 23.29\% \\
    DinoV2\cite{oquab2023Dinov2} + DPT\cite{ranftl2021vision} & Arxiv + ICCV 2021 & 2.858 & \cellcolor{bronze!25}3.421 & \cellcolor{silver!25}1.873 & 8.69\% & 20.24\% \\
    DepthPro\cite{bochkovskii2024depth} & Arxiv 2024 &2.920 & \cellcolor{silver!25}3.374 & \cellcolor{gold!25}1.677 & 12.46\% & 26.57\% \\
    MonodepthV2\cite{godard2019digging} & ICCV 2019 & 3.221 & 4.802 & 4.583 & 11.10\% & 25.55\% \\
    EndoSfMLearner\cite{ozyoruk2021endoslam} & MedIA 2021 & 2.661 & 3.926 & 3.878 & \cellcolor{silver!25}17.07\% & \cellcolor{silver!25}36.26\% \\
    AFSfMLearner\cite{shao2022self} & MedIA 2022 & \cellcolor{silver!25}2.461 & 3.459 & 3.440 & 13.20\% & \cellcolor{bronze!25}31.71\% \\
    IIDSfMLearner\cite{li2024image} & JHBI 2024 & \cellcolor{bronze!25} 2.516 & 3.435 & 3.408 & \cellcolor{bronze!25}15.76\% & 31.59\% \\
    EndoDAC\cite{cui2024endodac} & MICCAI 2024 & \cellcolor{gold!25}2.154 & \cellcolor{gold!25}3.141 & 2.281 & \cellcolor{gold!25}22.35\% & \cellcolor{gold!25}47.71\% \\
    \midrule
    MVS Methods & Source & MAE (mm) $\downarrow$ & RMSE (mm) $\downarrow$ & STD (mm) $\downarrow$ & $\delta < 1.1$ $\uparrow$ & $\delta_{1.25}$ $\uparrow$ \\
    \midrule
    
    MVSformer++\cite{cao2024mvsformer++} & ECCV 2024 & 2.405 & 4.934 &  4.296 & \cellcolor{gold!25}65.49\% & \cellcolor{gold!25}68.90\% \\
    Liu et.al\cite{liu2019dense} & IEEE TMI 2019 & \cellcolor{gold!25}1.343 &  \cellcolor{gold!25}2.235 & \cellcolor{gold!25}1.772 & 42.16\% & 60.60\% \\
    COLMAP\cite{schonberger2016structure} & CVPR 2016 & 1.588 & 2.529 & 1.947 &  35.12\% & 52.49\% \\
    \midrule
    Ours & Speed & MAE (mm) $\downarrow$ & RMSE (mm) $\downarrow$ & STD (mm) $\downarrow$ & $\delta < 1.1$ $\uparrow$ & $\delta_{1.25}$ $\uparrow$ \\
    \midrule
    Ours (iter 0) & \cellcolor{gold!75}3min/100frames & 0.1725 & 0.5343 & 0.5046 & 96.44\% & 99.83\% \\
    Ours (iter 1) & 5min/100frames & 0.1396 & 0.4693 & 0.4470 & 97.91\% & 99.87\% \\
    Ours (converge) & 10min/100frames & \cellcolor{gold!75}0.1252 & \cellcolor{gold!75}0.4427 & \cellcolor{gold!75}0.4238 & \cellcolor{gold!75}98.63\% & \cellcolor{gold!75}99.87\% \\
    \bottomrule
\end{tabular}
}

\caption{Quantitative comparison with state-of-the-art methods in synthetic endoscopy. In each category, \colorbox{gold!25}{gold}, \colorbox{silver!25}{silver}, and \colorbox{bronze!25}{bronze} indicate the 1st, 2nd, and 3rd best performance, respectively. MAE (mm), RMSE (mm), and STD (mm) denote the mean absolute, root mean square, and standard deviations of errors, respectively. The metrics $\delta < 1.1$ and $\delta_{1.25}$ represent the proportions of depths for which
$\max\!\Big(\tfrac{d_\text{pred}}{d_\text{gt}}, \tfrac{d_\text{gt}}{d_\text{pred}}\Big)$ 
is less than 1.1 or 1.25, indicating high-accuracy predictions.}
\label{tab:comparison}
\end{table}

\section{Experiment}

We evaluated our method through four experiments: \textbf{1) Virtual endoscopy:} Using simulated images and depth maps to assess point-to-point accuracy, which mainly tests the algorithm's potential, as acquiring ground truth depth is challenging in endoscopic scenes; \textbf{2) Phantom experiments:} Testing with real endoscopic data to evaluate iCT replacement potential; \textbf{3) Cadaver experiments:} Validating accuracy under cadaver conditions; and \textbf{4) Clinical intraoperative analysis:} Verifying stability with mucus and blood present, this part is focusing on qualitative assessment as direct quantitative patient testing and intraoperative CT were non-standard procedures. \textbf{The videos are shown on Supplementary Materials}

\textbf{Equipment included} a high-fidelity nasal phantom (FuseTec), HOPKINS 0° endoscope and IMAGE1 S 4U 4K camera (Karl Storz), with computation on an AMD 7950X CPU and dual RTX 4090 GPUs. Visual data was generated via VTK based on the same phantom. We obtained initial poses through COLMAP and employed selective stereo \cite{wang2024selective} for binocular depth estimation. Notably, our method's high structural accuracy made scale recovery both straightforward and accurate. We recovered the scale via ICP registration with CT models or between pre/postoperative reconstructions for iCT substitution tests.

\begin{figure}[tp]
    \centering
    \includegraphics[width=0.7\linewidth]{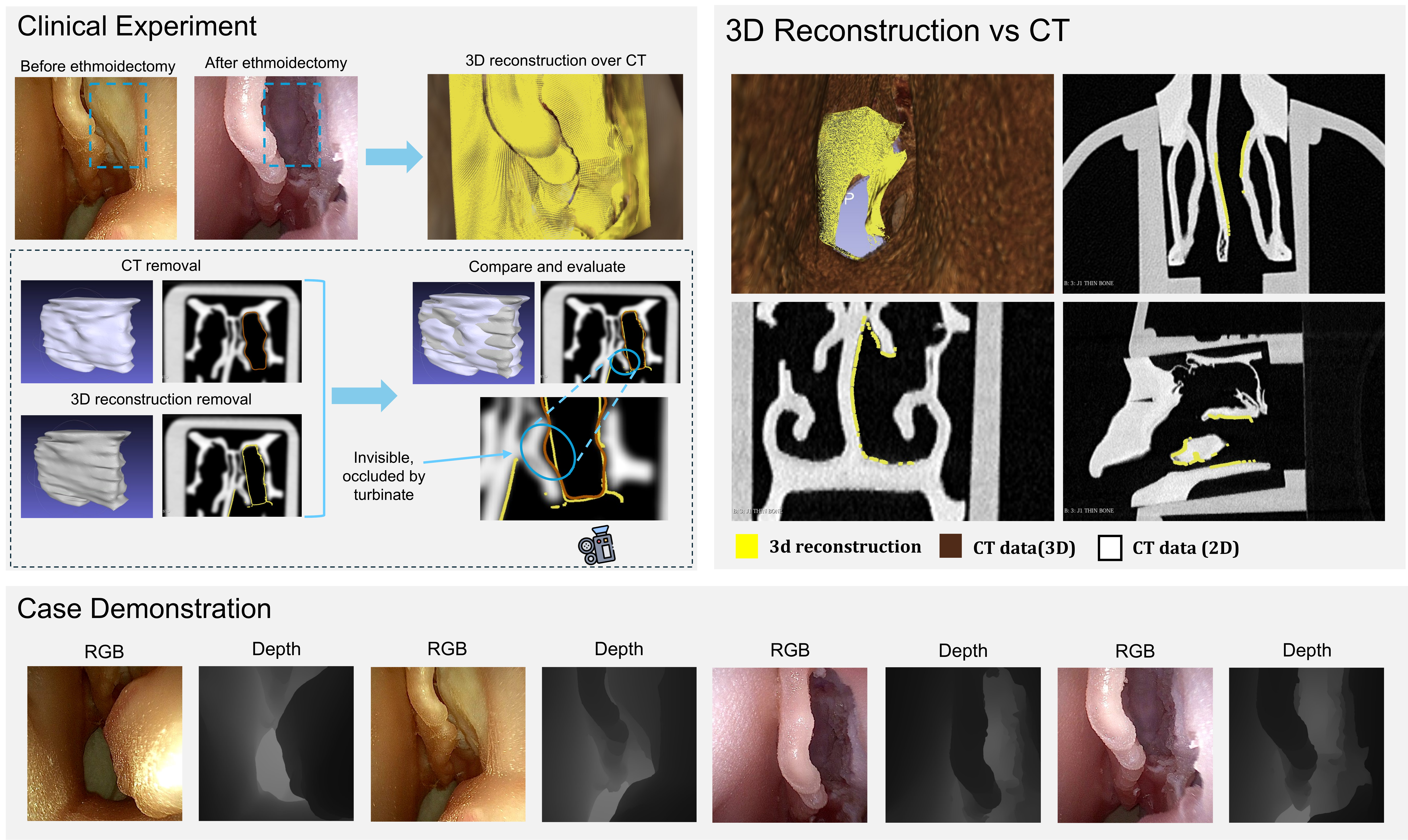}
    \caption{Comprehensive validation of our method across multiple experiments: \textbf{Clinical experiment} for validating its potential to substitute iCT; \textbf{3D Reconstruction vs. CT} shows near-perfect alignment along anatomical boundaries; \textbf{Case demonstrations} from endoscopic procedures show high structural fidelity of depth estimates with RGB inputs.}
    \label{fig:phantom}
\end{figure}

\subsubsection{Virtual Endoscopy Experiments:}

Table \ref{tab:comparison} presents our point-to-point depth evaluation results. Our method achieved superior accuracy (0.1252 ± 0.4427 mm in 10 min), surpassing all existing approaches and exceeding "one-slice-CT" accuracy at just 3-10 min/100 frames, significantly faster than iCT. MVS methods generally outperformed direct depth estimation approaches due to their joint optimization with camera poses, making them more robust to domain gaps between synthetic and real endoscopy where learning-based methods struggle to generalize. Among the models tested, medical-specific ones consistently outperformed general-purpose alternatives, with EndoDAC delivering the best performance in the direct depth estimation models.

\subsubsection{Phantom Experiments:}
For phantom experiments (Fig. \ref{fig:phantom}), we used iteration 1 results for time efficiency (5min end-to-end). An otolaryngologist performed an ethmoidectomy and endoscopically scanned the cavity, from which our method reconstructs the 3D surface. To evaluate iCT substitution potential, the otolaryngologist segmented both the post-operative CT cavity and our reconstruction. Results showed strong agreement (Table \ref{tab:quantitative_results}): reconstructed volume (2,936,mm\textsuperscript{3}) differed by only 4.05\% from ground truth (3,060,mm\textsuperscript{3}), with Hausdorff distance (0.2378,mm) and mean distance-to-mesh (0.2513,mm) confirming spatial accuracy. Discrepancies primarily occurred in endoscopically invisible regions (Fig. \ref{fig:phantom}, Clinical Experiment). In non-occluded areas (Fig. \ref{fig:phantom},3D Reconstruction VS CT), reconstruction closely matched CT, yielding mean distance-to-mesh of 0.2693,mm and Hausdorff distance of 0.2680,mm (Table \ref{tab:quantitative_results}, Phantom Geometric Analysis).

\subsubsection{Cadaver Experiments:}
We evaluated our method using cadaver specimens, with depth maps and 3D reconstructions shown in Fig. \ref{fig:cadaver-in vivo}A. Quantitative assessment was performed by comparing reconstructions against CT ground truth, with results presented in Table \ref{tab:quantitative_results}, Cadaver Geometric Analysis. Our method achieved a Mean Distance to Mesh of 0.2231 mm and a Hausdorff Distance of 0.2382 mm, closely matching the accuracy observed in phantom tests and demonstrating robust performance under realistic anatomical conditions.

\begin{table}
    \centering
    \resizebox{0.8\textwidth}{!}{
    \begin{tabular}{@{}lll@{}}
        \toprule
        \textbf{Category} & \textbf{Metric} & \textbf{Value} \\
       \midrule
        \multirow{2}{*}{Phantom Geometric Analysis} 
        & Mean Distance to Mesh & 0.2693 mm \\
        & Hausdorff Distance & 0.2680 mm \\
        \midrule
        \multirow{4}{*}{Phantom Volumetric Analysis} 
        & Reconstructed Volume & 2,936.0 mm³ \\
        & Ground Truth Volume (CT) & 3,060.0 mm³ \\
        & Volume Difference & 4.05\% \\
        & Hausdorff Distance & 0.2378 mm \\
        & Mean Distance to Mesh & 0.2513 mm \\
        \midrule
        \multirow{2}{*}{Cadaver Geometric Analysis} 
        & Mean Distance to Mesh & 0.2382 mm \\
        & Hausdorff Distance & 0.2231 mm \\        
        \bottomrule
    \end{tabular}
    }
    \vspace{0.5em}
    \caption{Quantitative evaluation results of our real endoscopic reconstruction method. The assessment includes: (1) phantom geometric comparison with CT mesh, (2) clinical volumetric analysis of the resected cavity performed by an otolaryngologist, and (3) cadaver geometric accuracy compared with CT mesh.} 
    \label{tab:quantitative_results}
\end{table}

\subsubsection{Clinical intraoperative data analysis:}
\begin{figure}[tp]
    \centering
    \includegraphics[width=0.96\linewidth]{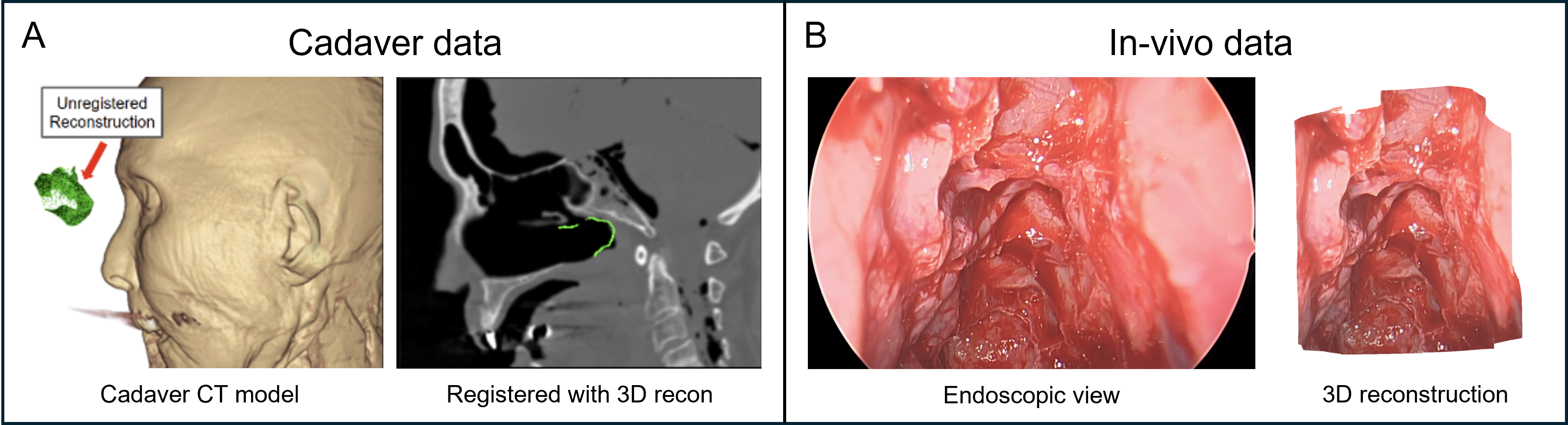}
    \caption{(\textbf{A}) Registration of the cadaver CT scan with our 3D reconstruction.
(\textbf{B}) Results on in vivo data, demonstrating that blood and mucus on the surface do not significantly affect reconstruction quality.}
    \label{fig:cadaver-in vivo}
\end{figure}

We further validated our method in clinical nasal surgeries under intraoperative conditions, as shown in Fig. \ref{fig:cadaver-in vivo} B. This evaluation focused on assessing robustness in the presence of surgical challenges such as blood and mucus. The resulting 3D reconstructions maintained high quality without significant artifacts, demonstrating the method's resilience to real-world endoscopic interference.

\section{Ablation study}

In this experiment, we evaluate the impact of optimization of novel camera pose generation on depth estimation accuracy. We used fixed baselines of 0.5 mm in synthetic endoscopy for iteration until convergence, and the results are shown in Table \ref{tab:Opt_comparison}. both groups show consistent improvement in point-to-point accuracy across iterations, demonstrating that even without perfect depth maps, the iterative refinement process effectively enhances reconstruction quality. This validates our iterative optimization approach. Additionally, while the \textbf{{\color{red!50!black}Without Opt}} group shows faster convergence in depth map accuracy, its best performance (0.180 mm MAE at convergence) remains inferior to even the worst performance of the \textbf{{\color{green!50!black}With Opt}} group (0.179 mm MAE at iteration 0). This conclusively demonstrates the effectiveness of our optimization strategy for novel view generation in achieving superior depth estimation results.

\begin{table}
    \centering
    \setlength{\tabcolsep}{4pt}
    \scriptsize
     \resizebox{0.7\textwidth}{!}{
    \begin{tabular}{@{}lclll@{}}
        \toprule
        \textbf{Iteration} & Baseline (mm) & MAE (mm) $\downarrow$ & RMSE (mm) $\downarrow$ & $\delta < 1.1$ $\uparrow$ \\
        \midrule
        \multirow{2}{*}{Iter 0} 
        & \multirow{2}{*}{\centering 0.5}
        & {\cellcolor{green!15}} 0.179 & {\cellcolor{green!15}} 0.547 & {\cellcolor{green!15}} 95.72\% \\
        & &{\cellcolor{red!15}} 0.285 &{\cellcolor{red!15}} 0.805 &{\cellcolor{red!15}} 90.57\% \\
        \midrule
        \multirow{2}{*}{Iter 1}
        & \multirow{2}{*}{\centering 0.5}
        & {\cellcolor{green!15}} 0.143 & {\cellcolor{green!15}} 0.471 & {\cellcolor{green!15}} 97.30\% \\
        & &{\cellcolor{red!15}} 0.201 &{\cellcolor{red!15}} 0.599 &{\cellcolor{red!15}} 93.53\% \\
        \midrule
        \multirow{2}{*}{Iter 2}
        & \multirow{2}{*}{\centering 0.5}
        & {\cellcolor{green!15}} 0.132 & {\cellcolor{green!15}} 0.457 & {\cellcolor{green!15}} 98.14\% \\
        & &{\cellcolor{red!15}} 0.187 &{\cellcolor{red!15}} 0.570 &{\cellcolor{red!15}} 94.00\% \\
        \midrule
        \multirow{2}{*}{Converge}
        & \multirow{2}{*}{\centering 0.5}
        & {\cellcolor{green!15}} 0.123 & {\cellcolor{green!15}} 0.434 & {\cellcolor{green!15}} 98.47\% \\
        & &{\cellcolor{red!15}} 0.180 &{\cellcolor{red!15}} 0.555 &{\cellcolor{red!15}} 94.54\% \\
        \bottomrule
    \end{tabular}
    }
    \caption{Comparison of depth estimation accuracy with and without novel view pose optimization across iterations. {\color{green!50!black}Green rows} represent results With Opt, {\color{red!50!black}Red rows} represent results Without Opt.}
    \label{tab:Opt_comparison}
\end{table}
\section{Conclusion}
This research presents a novel depth estimation and 3D reconstruction pipeline for monocular endoscopes in ESS, offering a promising alternative to iCT. In virtual endoscopy, it achieves “one-CT-slice” accuracy in 3 minutes per 100 frames, which is faster than iCT. Real endoscopy tests in phantom and cadaver confirm similar accuracy and speed, meeting clinical standards, the in-vivo test confirmed its robustness under complex surgical scenarios. However, the method is not yet real-time and is limited to confined spaces ($\leq$ 50 mm). Future work will focus on algorithmic and hardware optimizations, including integrating generalizable NeRF for near-real-time processing, as well as depth refinement for larger spaces, ultimately expanding clinical utility and improving outcomes in minimally invasive procedures.
\bibliographystyle{splncs04}
\bibliography{references}
%




\end{document}